\documentclass[
    ,final            
  ]
  {aipproc}

\layoutstyle{6x9}
\usepackage{hhline}
\usepackage{graphicx}
\usepackage{epsfig}
\usepackage{natbib}
\usepackage{amssymb}

\begin{document}

\title{Femtoscopy in heavy ion collisions:\\ Wherefore, whence, and whither?
  \thanks{Combined write-up of presentations at XXXV International Symposium on Multiparticle Dynamics, and at the
  Workshop on Particle Correlations and Femtoscopy, Krom\v{e}\v{r}\'{i}\v{z}, Czech Republic, August 2005}}


\author{Mike Lisa}{
  address={Physics Department, Ohio State University, 191 W. Woodruff Ave,, Columbus Ohio 43210, USA}
}

\begin{abstract}
  I present a brief overview of the wealth of femtoscopic measurements from the past
  two decades of heavy ion experiments.  Essentially every conceivable ``knob'' at
  our disposal has been turned; the response of two-particle correlations to these
  variations has revealed much about the space-momentum substructure of the hot source
  created in the collisions.  
  I discuss the present status of the femtoscopic program and questions
  which remain, and point to new efforts which aim to resolve them.
\end{abstract}

\maketitle

\begin{flushleft}
\begin{small}
    {\it The slowly crawling ants will eat our dreams.}\\ 
    Andre Bia\l as, musing on words of Andre Breton as they might apply to femtoscopy.\\
\medskip
    {\it Go to the ant, thou sluggard; consider her ways, and be wise.} - Proverbs vi.6
\end{small}
\end{flushleft}


\section{Wherefore}

High energy collisions between electrons, hadrons, or nuclei produce highly nontrivial systems.
Especially in the soft (low-$p_T$, long spatial scale) sector, the inclusive distributions of
the measured multiparticle final states are dominated by phase-space; to first order
the momentum spectra and particle yields appear thermal, revealing little of the underlying
physics of interest.  Detailed information in this sector is obtained only through correlations;
inclusive spectra tell much less than half the story.

In particular, multiparticle production is a {\it dynamic} process, evolving in space and time.
For several decades now, small relative momentum two-particle correlations have been used
to probe the space-time structure of systems at the femtometer scale.  Measurements
and constantly-improving techniques variously called ``intensity interferometry,''
``HBT,'' ``GGLP,'' ``non-identical correlations,'' etc, are nowadays discussed under
the common rubric of femtoscopy~\cite{Lednicky:2005af}, as the title of this new
workshop series reflects.

While understanding the space-time features of the system is important to both the
particle and the heavy ion physicist, in the latter case it is even vital.
After all, non-trivial geometrical effects {\it dominate} the physics of heavy ion collisions.

From the very broadest perspective, the entire heavy ion program is geared to
generate and study a qualitative change in the geometric substructure of the hot system.
Strongly-coupled~\cite{Gyulassy:2004zy} or not, the quark-gluon plasma (QGP) is
a soft QCD system, in which colored degrees of freedom are relevant over
large length scales.
Of particular interest is the existence and nature of a deconfinement phase transition;
a significant and sudden change in the degrees of freedom should be reflected in space-time
aspects of the system~\cite{Rischke:1996em}.
Also of generic importance is the (often unasked) question of whether the ``system''
generated is, indeed, a system.  Any discussion of ``matter'' or ``bulk'' properties
relies on an affirmative answer.

More specifically, geometry defines each stage of the system's evolution.
In the initial state, the entrance-channel geometry (impact parameter $\vec{b}$)
determines the subsequent collective evolution and anisotropic expansion of
the system~\cite{Huovinen:2001cy}; the resulting ``elliptic flow''~\cite{Ollitrault:1997vz}
has been the basis of $\sim25\%$ of the publications from the RHIC program.
In the intermediate state, also, geometry dominates: quantitative understanding
of exciting parton energy loss (or ``jet quenching'') measurements~\cite{Adams:2004wz,ColeQM05}
requires detailed information of the evolving size and anisotropic shape of the system.
If coalescence is indeed the mechanism of bulk hadronization~\cite{Fries:2003vb},
space-momentum correlations in the intermediate stage induce 
clustering
effects which must be modeled quantitatively~\cite{Molnar:2004zj}.

Clearly then, for the soft (bulk) sector in heavy ion collisions, geometrical issues
dominate both the physics of interest and the system with which it is probed.
No surprise, then, that since the relativistic heavy ion program began roughly two decades ago,
femtoscopic studies have played a major role, and a ``sub-community'' has developed.
It was not long before the erstwhile ``nuclear'' physicists contributed physical and
technical insights to a type of measurement initially borrowed from their particle physics colleagues.
At workshops like this, such dialogue continues unabated.

Several excellent reviews of femtoscopy in heavy ion physics
have very recently appeared in the literature~\cite{Lednicky:2005af,Csorgo:2005gd,Padula:2004ba,Lisa:2005dd}.
Together with physics discussions, the reader may find in them precise definitions of the correlation function,
``homogeneity lengths,''
``HBT radii,'' ``out-side-long'' coordinate system etc.
Here, I assume familiarity with such concepts, and very briefly review the status of heavy ion femtoscopy at present.
I emphasize the breadth of systematics which has been explored so far, what (we think) it has told us, and
what continues to puzzle us.  I then identify a 
few promising directions in which the field is moving, pointing for details on these to others' contributions
to these proceedings.

\section{Whence}

Due to their copious production and ease of detection,
most femtoscopic measurements have utilized correlations between charged pions.
Further, many experiments have focused on central ($|\vec{b}|=0$) collisions, since (1) azimuthal symmetry simplifies
the femtoscopic formalism~\cite{Lisa:2005dd,Heinz:2002au};
and (2) maximal energy densities and spatial extents are generated.
The extent of measured femtoscopic systematics 15-20 years ago is represented in Figure~\ref{fig:AlexanderBevalac},
showing that, in central collisions involving nuclei with mass number $A$, HBT radii scale approximately
as $A^{1/3}$~\cite{Bartke:1986mj,Alexander:2003ug}.  Apparently trivial, these data were at the same time
comforting, confirming that pion correlations did indeed track with geometric scales.

Since then, femtoscopic data and techniques have evolved tremendously,
generating an equally tremendous range of systematic femtoscopic studies.
The first femtoscopic measurement in truly relativistic heavy ion collisions was reported almost twenty years ago
by the NA35 Collaboration at the CERN SPS~\cite{Humanic:1988ny}.  
Similar measurements have been performed  at the SPS and the BNL AGS and RHIC accelerators
over the collision energy range $\sqrt{s_{NN}}\approx 2.3-200$~GeV.  Thus, in each of the two complementary quantities--
energy and time-- we may consider two decades' worth of systematics~\cite{Lisa:2005dd}.

The original hope was to find ``anomalously'' large
spatial and/or temporal scales, as reflected in the HBT radii, indicating large entropy generation or a long-lived
QGP state.  This expectation was considered rather generic~\cite{Harris:1996zx}, and, guided by quantitative predictions
from hydrodynamical models~\cite{Rischke:1996em,Bass:1999zq}, the most commonly-discussed systematic was the
excitation function (i.e. energy dependence) of pion HBT radii.
This is shown in Figure~\ref{fig:excitation}, where no striking features are observed
in the HBT radii at any collision energy.  As I discuss later, this observed contradiction of a seemingly-generic expectation may be
considered the second ``HBT puzzle.''

\begin{ltxfigure}[t!]
\begin{minipage}[t]{0.5\textwidth}
\centerline
{\includegraphics[width=1\textwidth]{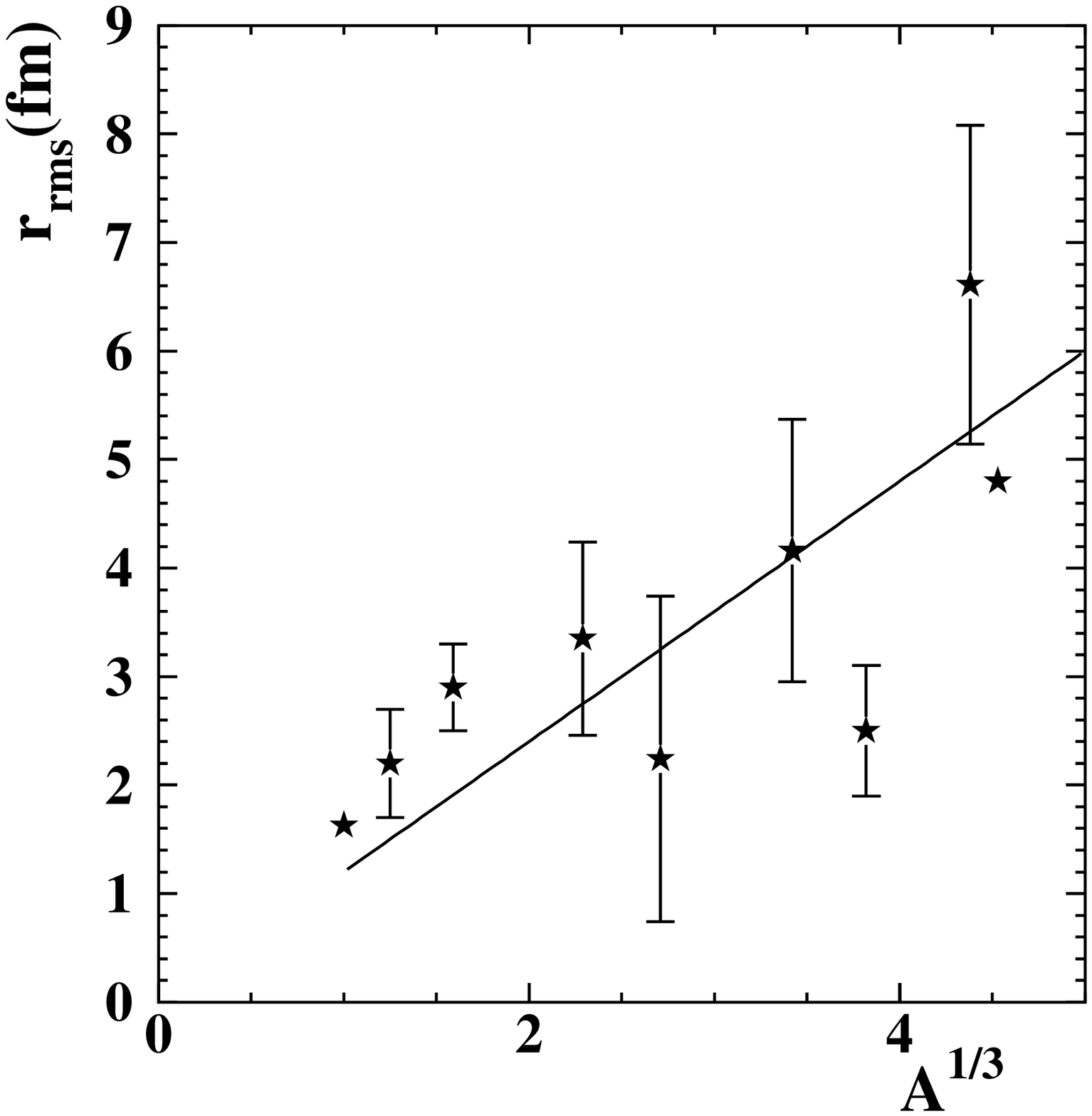}}
\caption{Pion HBT radius versus the mass number of colliding nuclei, from Bevalac experiments $\sim 20$ years ago.
                         Compilation from~\protect{\cite{Alexander:2003ug}}.
                         \label{fig:AlexanderBevalac}}
\end{minipage}
\hspace{\fill}
\begin{minipage}[t]{0.45\textwidth}
\centerline{\includegraphics[width=0.85\textwidth]{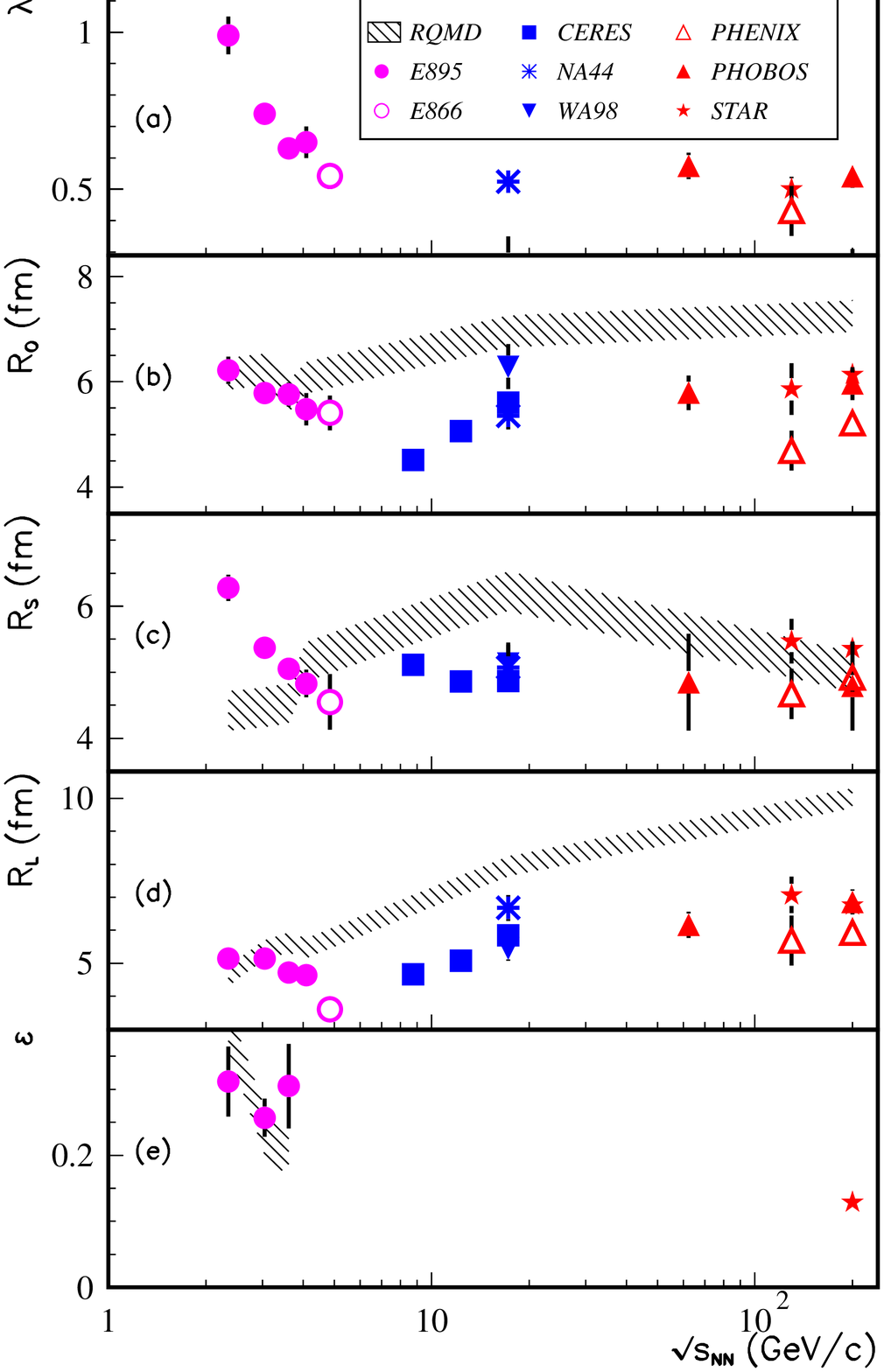}}
\caption{World dataset of published HBT radii from central Au+Au (Pb+Pb) collisions versus collision energy.
                          Compilation from~\protect{\cite{Lisa:2005dd}}.
                          \label{fig:excitation}}
\end{minipage}
\end{ltxfigure}

Clearly, insight into geometrically-driven physics requires more detailed systematic studies than the simple
excitation function.
Indeed, this has always been a generic requirement for extracting physics from {\it any} observable
in heavy ion physics, and has required
development of heavy ion programs with simultaneous, complementary, 
large-acceptance
experiments running at dedicated machines.  Especially in the crucial soft sector, more is learned
by varying independent variables than by long runs at the highest possible energy a given machine can deliver.

Inspired by recent ``schematic equations''~\cite{Gyulassy:2004jj},
I denote the impressive {\bf multi}-dimensional space explored by femtoscopic experiments as
\begin{equation}
\label{eq:multi-dim}
{\rm Heavy \ Ion \ Femtoscopy} = R\left(\sqrt{s_{NN}};A,B,|\vec{b}|,\phi,y,m_T,m_1,m_2\right)
\end{equation}

\noindent{\bf Global dependences}
Especially in light of ``puzzles,'' we need to perform a similar study as shown in Figure~\ref{fig:AlexanderBevalac},
checking that femtoscopic radii track with geometric collision scales to first order.
We may vary the geometric scale of the reaction zone by varying the atomic numbers of the colliding nuclei, $A$ and $B$,
and/or by selecting events of varying impact parameter, $|\vec{b}|$.
Of course, fixing only one of these parameters will not define the collision scale; instead, a natural 
quantity would be the number of participating nucleons $N_{part}$~\cite{Bialas:1976ed}.
Pion HBT radii corresponding to different $A$, $B$, $|\vec{b}|$ and $\sqrt{s_{NN}}$ are collected in
Figure~\ref{fig:Npart}.  The left panels show that these femtoscopic lengths scale similarly to
those shown in Figure~\ref{fig:AlexanderBevalac}, replacing $A$ by $N_{part}$.  (Note that results for central
collisions, $|\vec{b}|\approx 0$, are shown in Figure~\ref{fig:AlexanderBevalac}, so that $N_{part} \sim A$.)
The HBT radius $R_{out}$, which mixes space and time non-trivially, may be expected to violate a pure
geometrical scaling; this may explain the increased spread in the upper panels of Figure~\ref{fig:Npart}.

\begin{ltxfigure}[t!]
\begin{minipage}[t]{0.48\textwidth}
\centerline{\includegraphics[width=1.05\textwidth]{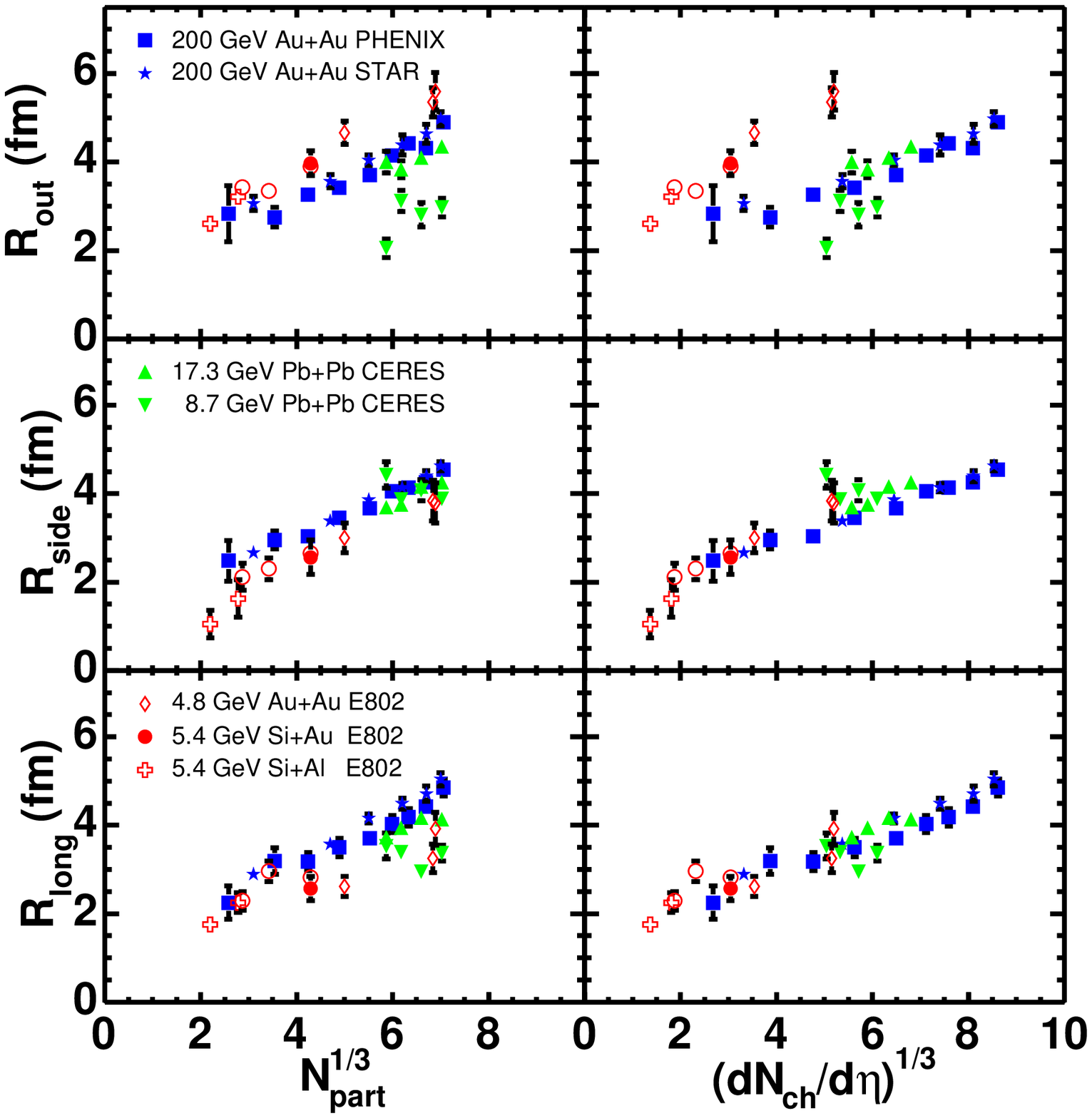}}
    \caption{Pion HBT radii plotted versus the number of participating nucleons (left panels), and versus the
                           charged particle multiplicity (right panels).  Compilation from~\protect{\cite{Lisa:2005dd}}.
                           \label{fig:Npart}}
\end{minipage}
\hspace{\fill}
\begin{minipage}[t]{0.48\textwidth}
\centerline{\includegraphics[width=1.05\textwidth]{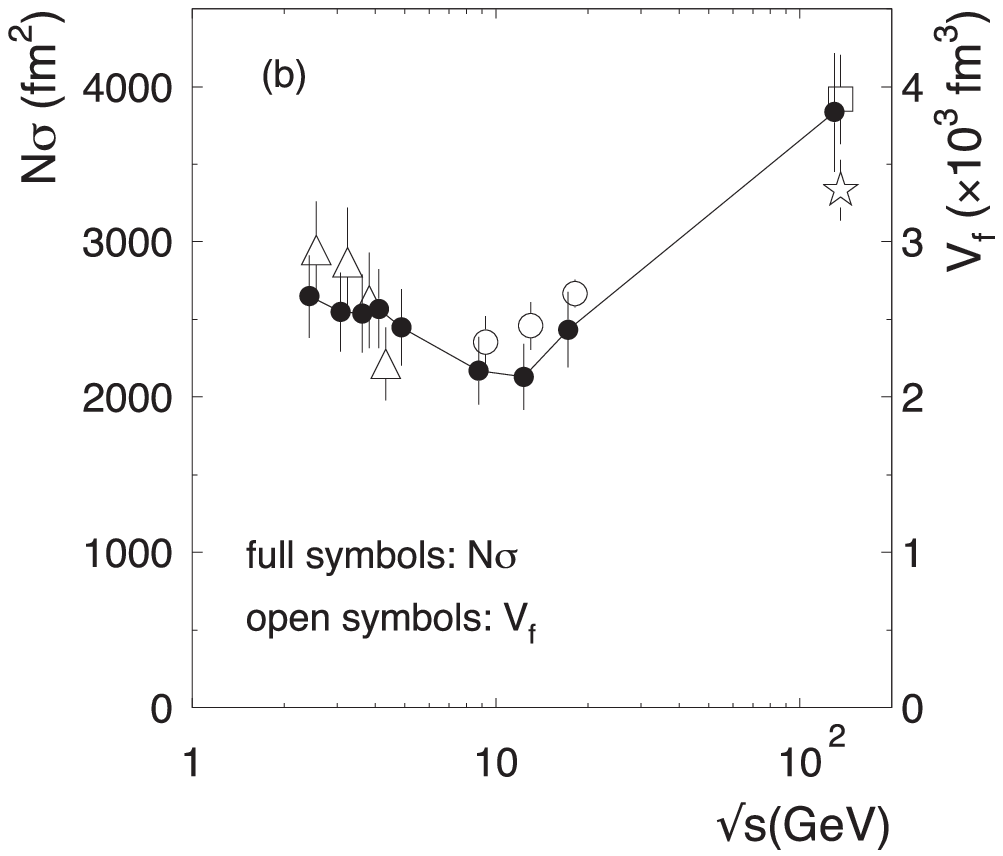}}
             \caption{The ``effective pion cross-section'' $N_{\rm proton}\cdot\sigma_{p\pi}+N_{\rm pion}\cdot\sigma_{\pi\pi}$
                      and the ``freezeout volume'' $\sim R_{long}\cdot R_{side}^2$ are plotted as a function of the collision energy,
                      for central Au+Au (Pb+Pb) collisions.  Figure and further details in~\protect{\cite{Adamova:2002ff}}.
\label{fig:CERESuniversal}}
\end{minipage}
\end{ltxfigure}

To good approximation, at a given $\sqrt{s_{NN}}$, total multiplicity (a final-state quantity) is a function only of $N_{part}$
(an entrance-channel quantity), independent of $A$, $B$, or $|\vec{b}|$.  The relationship does, however,
depend on collision energy~\cite{Back:2004je}.  As seen in the right panels of Figure~\ref{fig:Npart}, the final-state multiplicity
provides a more common scaling parameter than $N_{part}$; recent analyses~\cite{Adams:2004yc,Zibi:WPCF,Chajecki:2005zw} show that this scaling persists for
different $m_T$ values and for lighter colliding systems at RHIC.

Several observations may be made about this multiplicity scaling.  
Firstly, it appears that knowledge of $dN_{ch}/d\eta$ alone allows
``prediction'' of the HBT radii (at least $R_{long}$ and $R_{side}$).  This suggests that the small increase of these radii with $\sqrt{s_{NN}}$
seen in Figure~\ref{fig:excitation} is associated with increased particle production as the collision energy is raised.  (Note that $N_{part}$
is approximately constant for the data in Figure~\ref{fig:excitation}.)  
Secondly, the finite offset $d$ in the approximately linear
relationship $R_{long}\cdot R_{side}^2 = c\cdot (dN/d\eta) + d$ means that freeze-out does {\it not} occur at fixed density~\cite{Zibi:WPCF}.

Thirdly, the scaling shown in the figure breaks down dramatically for $\sqrt{s_{NN}} \lesssim 5$~GeV, as is obvious from the non-monotonic
behaviour seen in Figure~\ref{fig:excitation}.  As the CERES Collaboration has pointed out~\cite{Adamova:2002ff}, this is likely due to
the dominance of baryons at lower $\sqrt{s_{NN}}$.  Indeed, a quantitative connection between the number of protons and pions, and 
a product of HBT radii is possible, by assuming a universal ($\sqrt{s_{NN}}$-independent) mean free path at freezeout $\lambda_f$.
In Figure~\ref{fig:CERESuniversal}, the ``freezeout volume'' $\sim R_{long}\cdot R{side}^2$ and the
``effective pion cross-section'' $N_{\rm proton}\cdot\sigma_{p\pi}+N_{\rm pion}\cdot\sigma_{\pi\pi}$ are seen to coincide by scaling
the latter by $\lambda_f=1$~fm, apparently contradicting the standard assumption that freeze-out occurs when the mean free path becomes
much larger than the system size.

HBT radii and the ``freeze-out volume'' may be connected only the context of a model which includes dynamical effects like flow.
The analysis of~\cite{Adamova:2002ff} ignores such effects; however, its bottom line remains approximately valid, as flow
effects on HBT radii are expected to be small at low $p_T$~\cite{Retiere:2003kf}.

\noindent{\bf Kinematic dependences}
Insight on the dynamical evolution and geometric substructure of the emission region is gained by studying the
dependence of femtoscopic lengths on the next three parameters in ``Equation''~\ref{eq:multi-dim}.

\begin{ltxfigure}[t!]
\begin{minipage}[t]{0.48\textwidth}
\centerline{\includegraphics[width=1.05\textwidth]{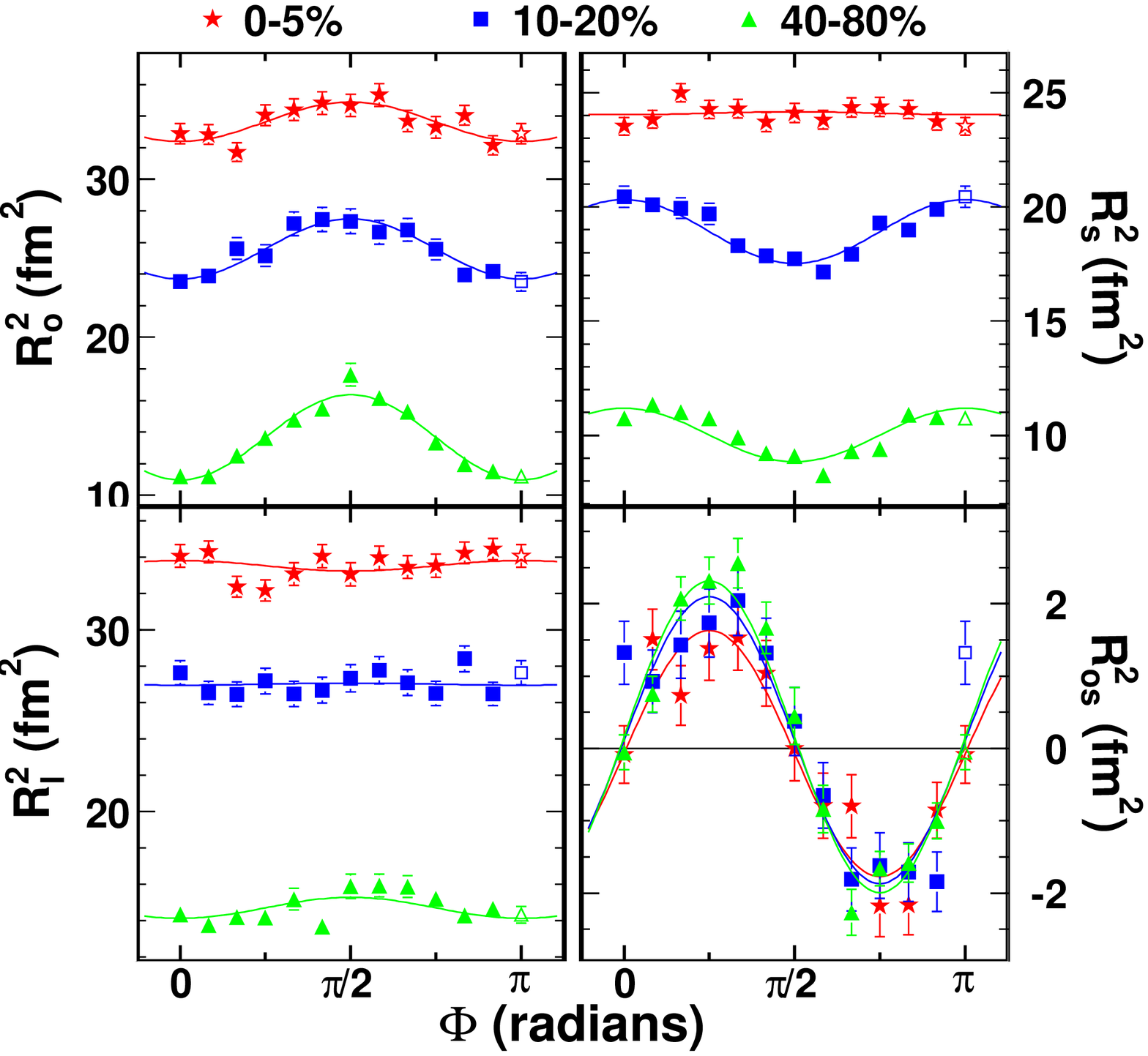}}
             \caption{Pion HBT radii measured for Au+Au collisions at $\sqrt{s_{NN}}$, plotted as a function
                      of azimuthal emission angle relative to the reaction plane.  From~\protect{\cite{Adams:2003ra}}
\label{fig:asHBT}}
\end{minipage}
\hspace{\fill}
\begin{minipage}[t]{0.48\textwidth}
\centerline{\includegraphics[width=1.05\textwidth]{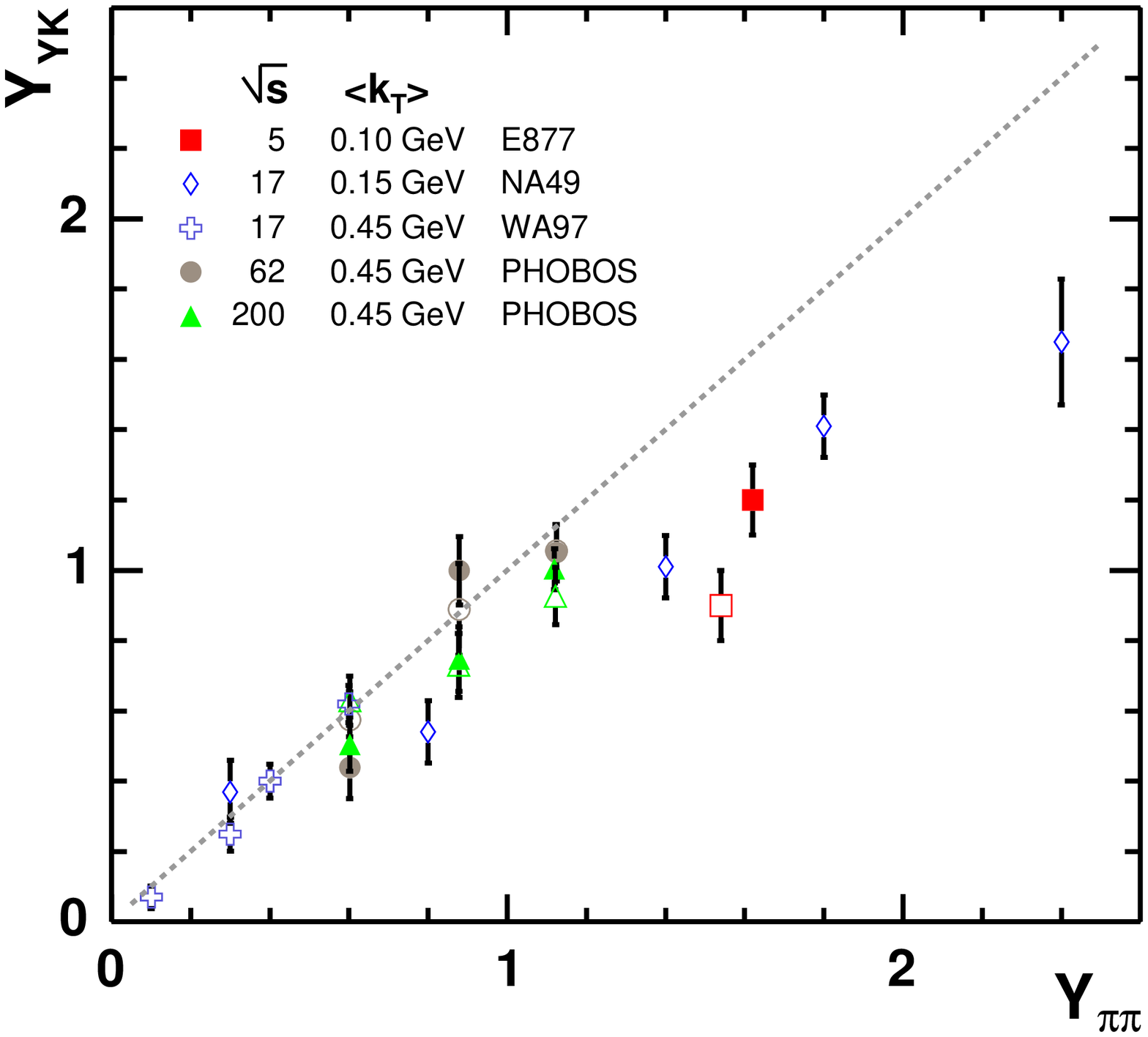}}
             \caption{The pion Yano-Koonin velocity (see text) versus pair rapidity for central Au+Au (Pb+Pb) collisions at
                      various energies
                            Compilation from~\protect{\cite{Lisa:2005dd}}.
\label{fig:Rapidity}}
\end{minipage}
\end{ltxfigure}

In non-central collisions, the entrance-channel geometry is naturally anisotropic; the hot source geometry approximates the
overlap between target and projectile, and is characterized by a ``long axis'' perpendicular to the impact parameter vector $\vec{b}$.
At RHIC, the system expands more rapidly in-plane ($\parallel \vec{b}$) than out ($\perp \vec{b}$)~\cite{Adams:2003zg}.  If it is,
indeed a collective {\it system} with finite lifetime, then the overall shape should evolve.  Pion HBT radii have been measured as
a function of their azimuthal angle $\phi_{\rm pair}\equiv\angle\left(\vec{K},\vec{b}\right)$ for Au+Au collisions.  The measurement
at RHIC~\cite{Adams:2003ra} is shown in Figure~\ref{fig:asHBT}.  There, it is clear that as $|\vec{b}|\rightarrow 0$, the freezeout source
becomes larger and rounder.  In fact, there is a nice ``rule of two''-- the source expands to twice its original size~\cite{Adams:2004yc,Zibi:WPCF}, and its
anisotropy $\epsilon\equiv\left(\langle y^2\rangle - \langle x^2\rangle\right)/\left(\langle y^2\rangle + \langle x^2\rangle\right)$
decreases by the same factor~\cite{Adams:2003ra}.  The relatively small change in the source shape 
is at least semi-quantitatively~\cite{Lisa:2003ze} consistent
with short timescale estimates~\cite{Retiere:2003kf} based on the longitudinal radius, and at variance
with expectations from ``realistic'' simulations~\cite{Teaney:2001av}.

As will become increasingly clear, the only femtoscopic systematic which might display non-trivial $\sqrt{s_{NN}}$ dependence is, in fact, the dependence on $\phi_p$.
This is clear from the bottom panel of Figure~\ref{fig:excitation}, in which the relative paucity of such measurements is also clear.
It will be especially interesting to see whether the flow and/or timescales
at the LHC are sufficiently large to produce in-plane freeze-out configurations~\cite{Heinz:2002sq}.

Experiments at a wide range of collision energies have mapped out the rapidity dependence of pion HBT radii.  Of particular interest here is
the so-called Yano-Koonin rapidity $Y_{\rm YK}$~\cite{Yano:1978gk,Wu:1996wk}, which should approximate the rapidity of the fluid element which emits a
pair of pions at some rapidity $Y_{\pi\pi}$.  Figure~\ref{fig:Rapidity} shows an approximately ``universal'' behaviour $Y_{\rm YK}\approx Y_{\pi\pi}$,
independent of $\sqrt{s_{NN}}$.  This is consistent with (but not proof of~\cite{Lisa:2005dd}) emission from a boost-invariant system~\cite{Wu:1996wk}.

\begin{figure}[t!]
\centerline{\includegraphics[width=1\textwidth]{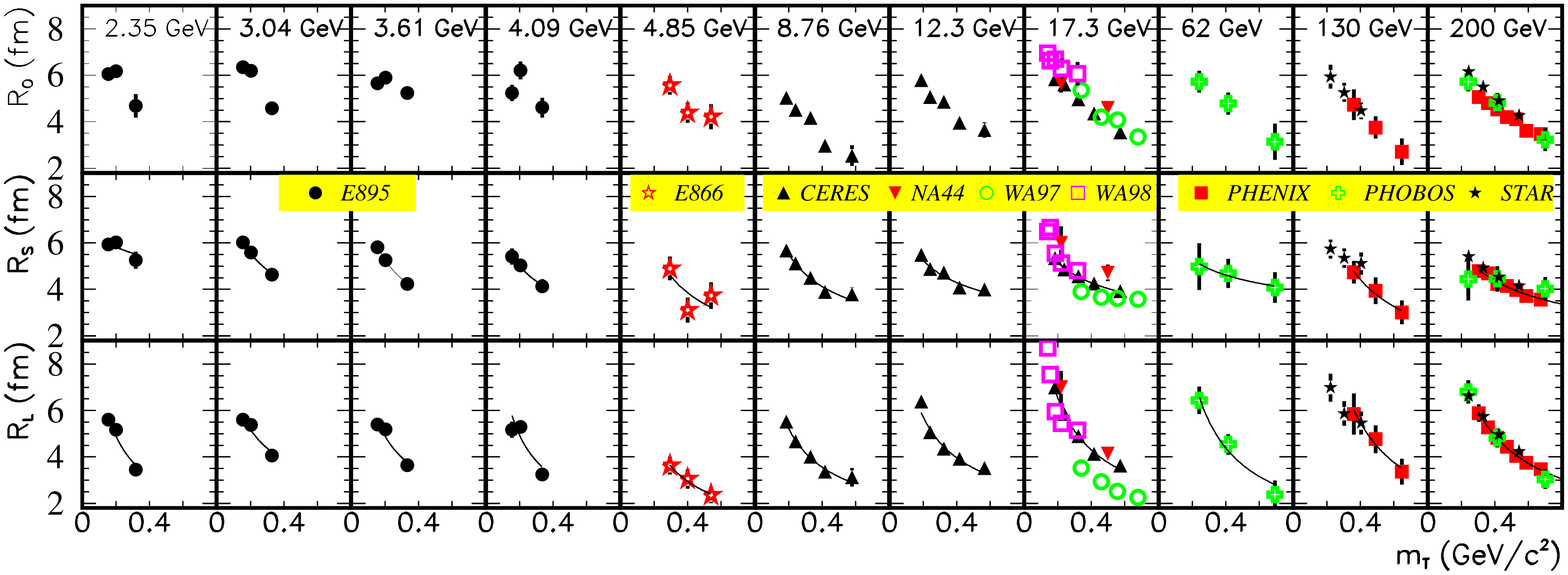}}
    \caption{Pion HBT radii plotted versus the transverse mass $m_T$ for all published measurements of central
             Au+Au (Pb+Pb) collisions over two decades in $\sqrt{s_{NN}}$.  Compilation from~\protect{\cite{Lisa:2005dd}}.
                           \label{fig:pT}}
\end{figure}

The most extensively-studied kinematic systematic has been the $p_T$-dependence of pion HBT parameters.
Figure~\ref{fig:pT} shows the world dataset of published measurements for central Au+Au (Pb+Pb) collisions.
The falling dependence of femtoscopic scales on transverse velocity is generally believed to arise from
collective transverse and longitudinal flow (e.g.~\cite{Retiere:2003kf}).  As I mentioned earlier, strong
collective flow would be an indication that a real {\it bulk} system has been formed.  

\begin{ltxfigure}[t!]
\begin{minipage}[t]{0.4\textwidth}
\centerline{\includegraphics[width=0.8\textwidth]{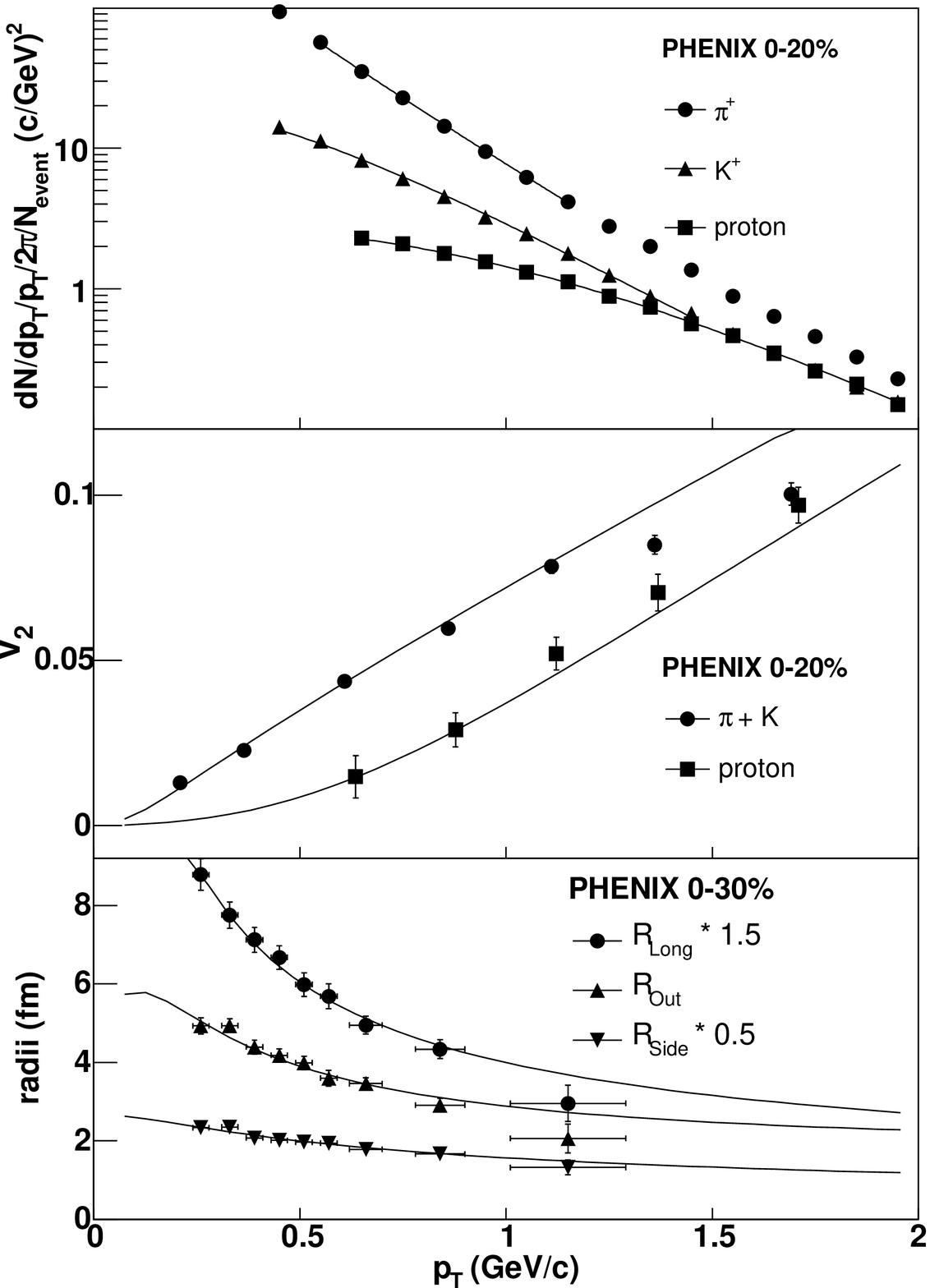}}
    \caption{A blast-wave~\cite{Retiere:2003kf} fit reproduces several observables at RHIC.  See text for details.
            From~\protect{\cite{Retiere:2004wa}}.
                           \label{fig:FabriceQM04}}
\end{minipage}
\hspace{\fill}
\begin{minipage}[t]{0.55\textwidth}
\centerline{\includegraphics[width=1.2\textwidth]{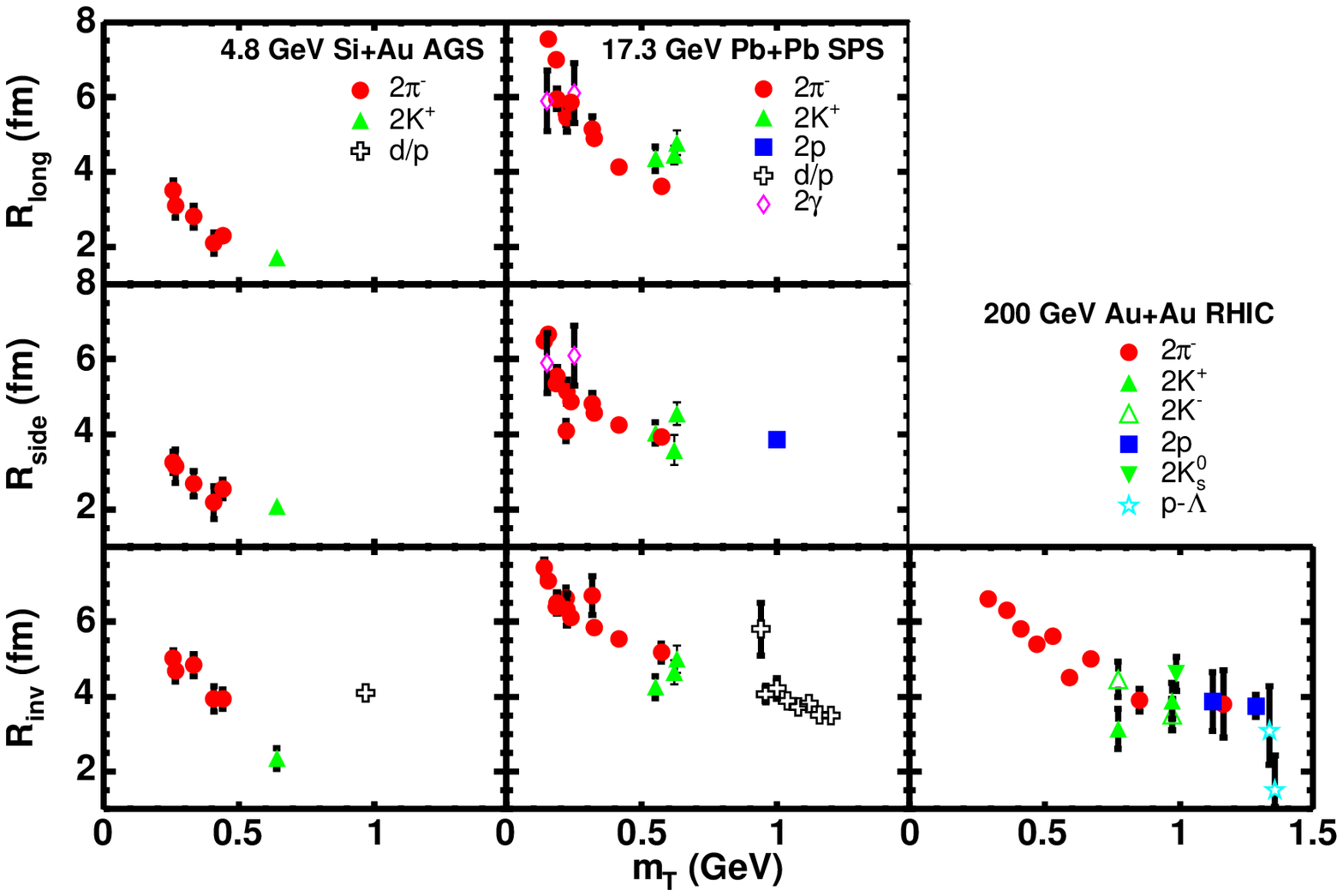}}
    \caption{Femtoscopic radii for various similar-mass particle pairs, plotted as a function of $m_T$.
             Compilation from~\protect{\cite{Lisa:2005dd}}.
                           \label{fig:mT}}
\end{minipage}
\end{ltxfigure}

The longitudinal radius scales approximately as $R_l \sim m_T^{-0.5}$, indicating strong longitudinal flow and
again consistent with expectations for emission from
a boost-invariant system~\cite{Akkelin:1995gh,Retiere:2003kf}.  Decreasing transverse radii $R_o$ and $R_s$ may be due
to collective transverse flow.
The simplest flow-dominated models quantitatively interrelate these femtoscopic $m_T$ dependences with other observations.
An example is shown in Figure~\ref{fig:FabriceQM04}, in which a very simple
freeze-out scenario~\cite{Retiere:2003kf}-- thermal motion superimposed on a collectively exploding source--
can simultaneously describe a broad range of data measured at RHIC.  The momentum-space distribution, quantified
by the average number distribution (top panel of Figure~\ref{fig:FabriceQM04}) and the number variation as a function
of azimuthal angle (middle panel) give an incomplete picture by themselves.  Momentum-dependent femtoscopic radii 
(bottom panel) probe the dynamical {\it sub}-structure of the collision, constraining models 
more stringently~\cite{Lisa:2005dd,Retiere:2003kf,Tomasik:2005ny,Chajecki:2005qm}.

\noindent{\bf Particle-species dependences}
Within the past several years, high-statistics datasets in experiments with good particle identification have allowed
the mapping of femtoscopic systematics with the final variables
in ``Equation''~\ref{eq:multi-dim}-- the mass (or species) of the correlated particles.

Signals of a {\it system's} collectivity at freeze-out should not be limited to the pions.  In the simplest picture, corresponding
to flow-dominated models (e.g.~\cite{Retiere:2003kf}) of Figure~\ref{fig:FabriceQM04}, femtoscopic radii should approximately scale with $m_T$, independent of
particle type.
An impressive common scaling of radii from {\it all} measured particles is, indeed, observed at {\it all} energies explored, as seen in Figure~\ref{fig:mT}.
The common scaling is particularly striking when one considers the quite different measurement systematics involved in charged pion correlations and,
say, $K^0_s$ correlations.  Even generalized nucleon separation scales, probed by relative yields of deuterons and protons ($d/p$ in Figure~\ref{fig:mT}),
follow the systematic, with the exception of one outlier point at the lowest energies.

\begin{table}[t]
\centering{
\begin{tabular}{|p{4mm}||p{16mm}|p{16mm}|p{5mm}|p{5mm}|p{5mm}|p{11mm}|p{7mm}|p{4mm}|p{4mm}|p{4mm}|p{4mm}|}
\hhline{------------}
~ & $\pi^+$ & $\pi^-$ & $K^+$ & $K^-$ & $K^0_s$ & $p$ & $\bar{p}$ & $\Lambda$ & $\bar{\Lambda}$ & $\Xi$ & $\bar{\Xi}$\\
\hhline{============}
$\bar{\Xi}$       & \cite{Chaloupka:WPCF}   & \cite{Chaloupka:WPCF}   &    &    &    &    &    &    &    &    &   \\ \hhline{------------}
$\Xi$             & \cite{Chaloupka:WPCF}   & \cite{Chaloupka:WPCF}   &    &    &    &    &    &    &    &    \\  \hhline{-----------}
$\bar{\Lambda}$   &    &    &    &    &    &  \cite{Adams:2005ws}  & \cite{Adams:2005ws}   &    &    \\  \hhline{----------}
$\Lambda$         &    &    &    &    &    &  \cite{Adams:2005ws}  & \cite{Adams:2005ws}   &    \\  \hhline{---------}
$\bar{p}$         & \cite{Adam:WPCF}   & \cite{Adam:WPCF}   & \cite{Adam:WPCF}   & \cite{Adam:WPCF}   &    & \cite{Hanna:WPCF}   & \cite{Hanna:WPCF}   \\  \hhline{--------}
$p$               & \cite{Adam:WPCF}   & \cite{Adam:WPCF}   & \cite{Adam:WPCF}   & \cite{Adam:WPCF}   &    & \cite{Hanna:WPCF,Heffner:2004js}   \\  \hhline{-------}
$K^0_s$           &    &    &    &    &  \cite{Bekele:2004ci}  \\  \hhline{------}
$K^-$             & \cite{Adams:2003qa}   & \cite{Adams:2003qa}   &    & \cite{Heffner:2004js}   \\  \hhline{-----}
$K^+$             & \cite{Adams:2003qa}   & \cite{Adams:2003qa}   & \cite{Heffner:2004js}   \\  \hhline{----}
$\pi^-$           & \cite{Adams:2004yc,Hanna:WPCF}   &   \cite{Adams:2004yc,Adler:2004rq,Back:2004ug} \\  \hhline{---}
$\pi^+$           & \cite{Adams:2004yc,Adler:2004rq,Back:2004ug}   \\  \hhline{--}

\end{tabular}
}        
\caption{A very incomplete table of published or ongoing femtoscopic studies at RHIC for various particle combinations.
         ``Traditional'' identical-particle interferometry lies along the lowest diagonal line of cells.
\label{tab:nonid}}
\end{table}

Correlations between non-identical particles probe not only the sizes, but also the 
relative displacement of the particles' emission zones in space-time~\cite{Lednicky:2001qv}.
Any collective freeze-out scenario naturally implies a specific relationship between emission regions of the various particle types.
In a flow-dominated picture, the emission zones for high-$m_T$ particles are not only smaller than those for low-$m_T$ particles,
but are also inevitably located further from the center of the collision region~\cite{Retiere:2003kf}, as suggested by the schematic in
Figure~\ref{fig:AnnRevFig03}.
As discussed in detail in the contribution of A. Kisiel~\cite{Adam:WPCF}, available measurements of these displacements at RHIC
provide further support of the flow-dominated freeze-out scenario.

Non-identical particle correlations are today a growth industry.  Table~\ref{tab:nonid} lists only a sampling of recently-published
or ongoing analyses at RHIC energies.  Similar studies have been performed at lower energies~\cite{Lednicky:WPCF,Lisa:2005dd}.
The diagonal axis corresponds to identical-particle correlations, the ``traditional'' focus of HBT interferometry.

\begin{ltxfigure}[t!]
\begin{minipage}[t]{0.48\textwidth}
\centerline{\includegraphics[width=1.15\textwidth]{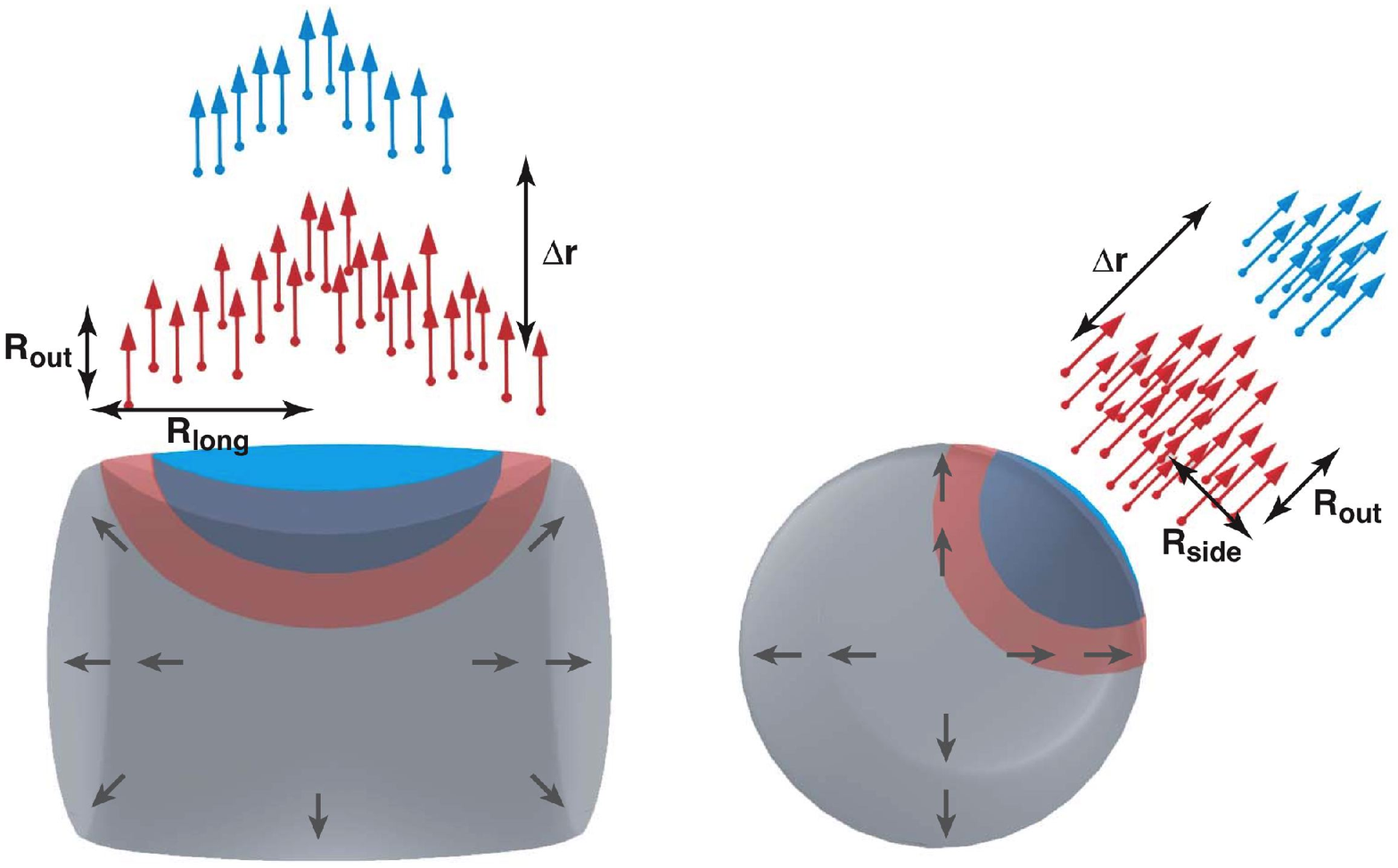}}
    \caption{Freezeout regions for particles of different species (or different transverse masses) emitted from a common source.
             Two-particle correlations measure the (momentum-dependent) size, shape, and orientation of the emission regions, as well as the
             average displacement ($\Delta r$) in the outward direction.  From~\protect{\cite{Lisa:2005dd}}.
                           \label{fig:AnnRevFig03}}
\end{minipage}
\hspace{\fill}
\begin{minipage}[t]{0.48\textwidth}
\centerline{\includegraphics[width=1.15\textwidth]{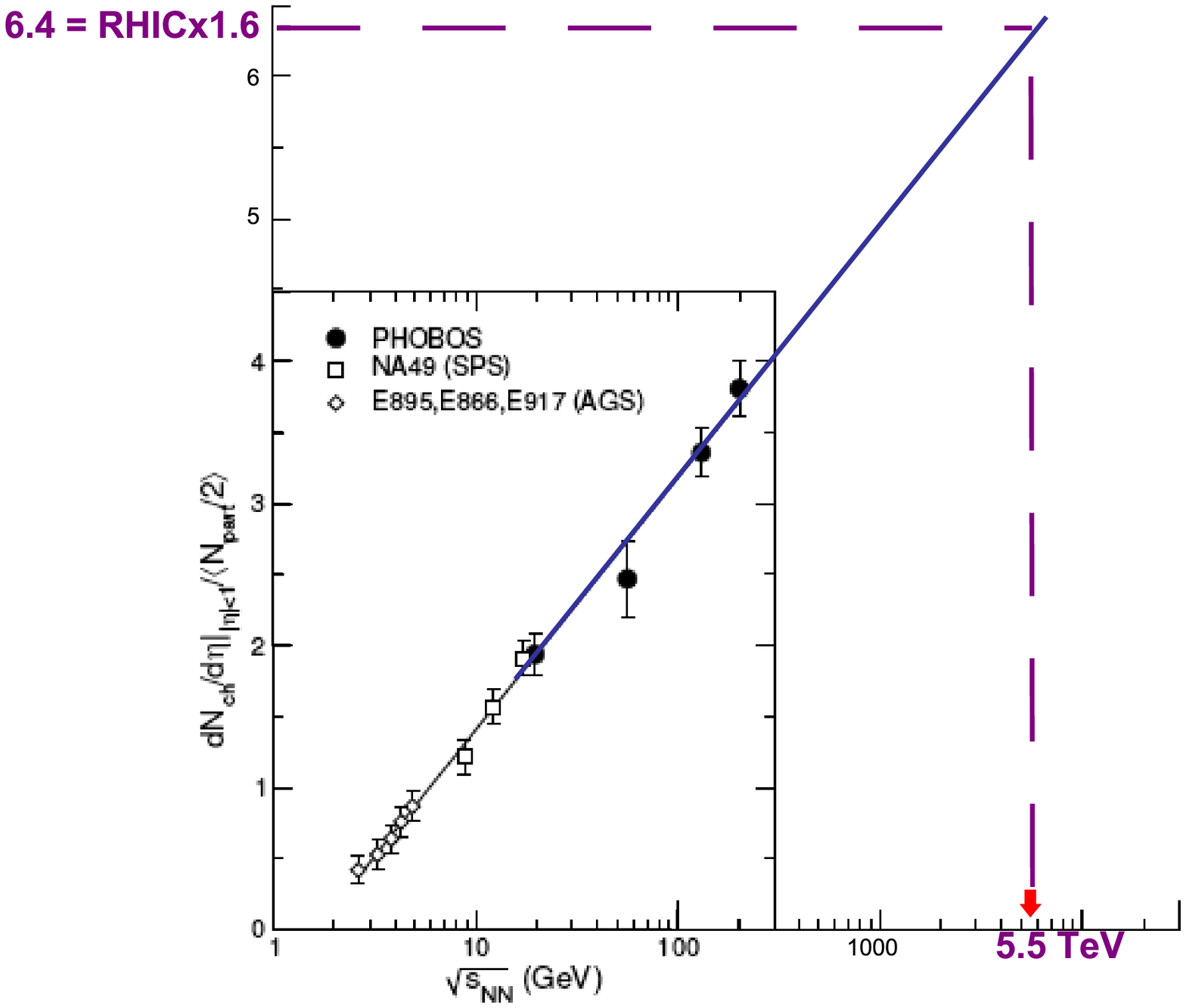}}
    \caption{Multiplicity density per participant pair, measured for Au+Au (Pb+Pb) collisions at AGS, SPS, and RHIC
             are shown in the lower panel, taken from~\protect{\cite{Back:2004je}}, and naively extrapolated by the author to LHC energy.
                           \label{fig:LHCextrap}}
\end{minipage}
\end{ltxfigure}

\section{Whither}

Excluding aficionados attending workshops such as this one, in the minds of most heavy ion
(or high energy) physicists, the term ``femtoscopy'' (or, more likely ``HBT'') brings to mind
only the single, rather uninspiring systematic plotted in Figure~\ref{fig:excitation}; indeed,
some may be tempted toward the dismissive view that
``the measured radius is always 5 fm.''

As we have just discussed, this is grossly unfair: the systematics are tremendously richer, with femtoscopic length scales
varying with almost every parameter in ``Equation''~\ref{eq:multi-dim}.
Furthermore, these strong systematic trends are found consistently by experiments separated by decades
and using quite different measurement and correction techniques; indeed, especially at RHIC, the data are
almost embarrassingly consistent (c.f. Figure~\ref{fig:pT}).
Yet further, it appears that these systematics may be well understood in the commonly-accepted
framework of {\it system} evolution due to strong flow quantitatively consistent with momentum-space
observables~\cite{Retiere:2003kf}.
Clearly, there is much more to femtoscopy than its most notorious Figure.

On the other hand, the trends shown in Figures~\ref{fig:Npart}, \ref{fig:Rapidity}, \ref{fig:pT}, and~\ref{fig:mT}
suggest that the notorious Figure~\ref{fig:excitation} quite correctly summarizes the situation after all.
At any $\sqrt{s_{NN}}$, the systematics of ``Equation''~\ref{eq:multi-dim} are quite rich and may well be
reconciled with a reasonable physical picture.  However, in more ways than expressed by Figure~\ref{fig:excitation},
those systematics are essentially independent of $\sqrt{s_{NN}}$!
Without resorting to agreement or disagreement with particular models, this second~\footnote{
In this experimental overview, I have not discussed what has come to be known as ``the'' HBT puzzle~\cite{Heinz:2002un}
which, simply put, is that otherwise-successful and apparently reasonable models like hydrodynamics do not reproduce
femtoscopic measurements~\cite{Lisa:2005dd}.  To all but the novice heavy ion physicist, however, the initial failure
of dynamical models to reproduce {\it diverse} observations is hardly puzzling.  The experience at lower energies is
that such initial failure is more the rule than the exception.  In light of other, more generic puzzles, I call this
problem only the ``{\it first} HBT puzzle''~\cite{Chajecki:2005qm}.
}
femtoscopic puzzle is startling, suggesting that the space-time consequences of the physical processes are the
same at RHIC as they are near the pion production threshold.  
Often, ``universal'' behaviour is a key to deeper physical insight.  Heavy ion femtoscopy,
however, might display a bit {\it too} much universality.

\noindent{\bf Whither... or wither?}

Given the prominence of nontrivial geometry to
the physics of heavy ion collisions in general, and the rather generic~\cite{Harris:1996zx} expectation of significant
changes in spacetime evolution with $\sqrt{s_{NN}}$, understanding this universality remains urgent.  What future efforts
might shed some light?

Today, one almost reflexively points to the impending heavy ion program at the LHC for new observations
generating fresh insights.  While anything might happen in an unexplored energy domain, we may venture a prediction.
Instead of a crystal ball, however, we use a mirror to gaze over our shoulder at twenty years of systematics in heavy ions.
The $\sqrt{s_{NN}}$-dependence of the global multiplicity (per participant pair) has been significantly extended at RHIC
and summarized by the PHOBOS collaboration~\cite{Back:2004je}.  Boldly and probably naively extending this systematic
leads to the expectation that multiplicities at the LHC will be $\sim 60\%$ higher than they are at RHIC,
as shown in Figure~\ref{fig:LHCextrap}.

As discussed in the previous Section, femtoscopic length scales--
for any $m_T$, $y$, $N_{part}$, or particle species-- depend primarily on event multiplicity.  
Taken together, Figures~\ref{fig:LHCextrap} and~\ref{fig:Npart} suggest that radii\footnote{
Precise expectations $R_{out}$  or $\phi$-dependent radii at the LHC  are, admittedly, less certain, as the former does not
scale exactly with multiplicity (cf Figure~\ref{fig:Npart}), and the multiplicity-dependence of the latter has not
been extensively mapped (cf Figures~\ref{fig:excitation} and~\ref{fig:asHBT}).
} in central collisions at
RHIC will simply be $\sim 17\%$ higher than they are at RHIC 
($(1.6)^{1/3} = 1.17$.).


Notably, evidence is mounting that perhaps {\it all} soft-sector 
observables are determined primarily by total multipilcity, independent of $\sqrt{s_{NN}}$.  Properly-scaled
elliptic flow~\cite{Roland:2005ei} and even strangeness enhancement~\cite{caines:QM05} appear to show universal multiplicity scalings.
Whether this is a trivial implication of entropy-driven phasespace dominance in observables sensitive to bulk medium is unclear.
However, nontrivial new phases of {\it matter} should have signatures in the long-distance (soft momentum) sector;
dependence only on multiplicity and not reaction energy would be intriguing.

So, perhaps the choice of collider facility (LHC versus RHIC) is unimportant,
and heavy ion femtoscopists should focus on filling in the holes of Table~\ref{tab:nonid}?
Most evidence thus far indicates that flow-dominated freezeout scenarios (e.g.~\cite{Retiere:2003kf}) fitted to identical $\pi$ correlations
essentially ``predict'' femtoscopic data using other particle combinations.  The data is yet scant, however, so ongoing
studies~\cite{Adam:WPCF} to further explore this Table are quite important.
There are even preliminary reports~\cite{Chaloupka:WPCF}, with exotic particle combinations, of inconsistencies with these freezeout models.
If confirmed, strong theoretical focus should come to bear on this result.  If, on the other hand, varying the particle combination
repeatedly yields results ``predicted'' by blast-wave models, continually filling in cells of Table~\ref{tab:nonid} risks becoming
a stamp-collecting exercise.

The most important recent experimental developments in heavy ion femtoscopy were presented at these workshops.  In addition to those
discussed above, I briefly mention here some ongoing studies which I find most promising.

Even if all of the particle combinations in Table~\ref{tab:nonid} follow simple blast-wave calculations, and so reveal no
new femtoscopic information, this can actually be turned to good use.  In particular, one may turn around the traditional approach
in which one uses the known the two-particle final state interaction (FSI) to extract geometric information, to extract the FSI
itself~\cite{Lednicky:WPCF,Retiere:WPCF}.  Finalized results from STAR on $p-\Lambda$ correlations~\cite{Adams:2005ws} have extracted
previously inaccessible phase shift information for low-energy baryon-antibaryon scattering.  While not QGP-related physics, such
studies can make a unique contribution to low-energy QCD and hadronic physics.

Ideally suited workshops like these are studies which directly compare for the first time, at a fixed energy and using identical detector and analysis
techniques, correlation data from the heaviest ion collisions to that from p+p collisions~\cite{Zibi:WPCF}.  As has been observed
previously in high energy experiments, femtoscopic radii from identical pion correlations measured in p+p collisions decrease with
increasing $p_T$, qualitatively similar to the dependence shown in Figure~\ref{fig:pT}.  The preliminary STAR data shows, however,
that in all three HBT radii, the $p_T$ dependence is {\it quantitatively identical} in p+p and A+A collisions!  Since the heavy ion and
high energy communities have traditionally used very different physics mechanisms to explain this dependence, this observation potentially
throws the explanations of both communities into doubt.  If this result is confirmed, it ranks as the ``third HBT puzzle''~\cite{Chajecki:2005qm}.

Unexplained long-range structure in the correlation functions for the lowest-multiplicity collisions, however, presently cloud the interpretation
of the HBT radii for p+p collisions~\cite{Zibi:WPCF}.  Partly in an effort to understand this, a new representation of the data in terms of spherical
harmonic amplitudes in $\vec{q}$-space was developed as an experimental diagnostic tool~\cite{Chajecki:2005qm}.
In fact, a similar harmonic decomposition method was earlier already developed by Danielewicz and
collaborators~\cite{Danielewicz:2005qh,Danielewicz:WPCF}, not as a diagnostic,
but as a direct link to the detailed geometry (beyond simply length scales) of the emitting source.  This representation has a natural connection
to source imaging~\cite{Brown:WPCF}, and, indeed, first applications to PHENIX data have been reported~\cite{Chung:WPCF}.

Harmonic decompositions as an improved representation of the correlation function and source imaging as an improved, generalized fit to the data
are, in a sense, merely technical improvements, but they are quite significant ones.  Just as femtoscopic studies have explored the systematic landscape
of Equation~\ref{eq:multi-dim}, so should they probe the ``microscape'' of fine details of the measured data.

The femtoscopy of heavy ion collisions can be an addicting endevour.  Systematics make sufficient sense that we are convinced that
we are probe geometry at the femtometer scale.  Spacetime geometry at that scale is sufficiently important to the physics that the
measurements must be done well.  Such measurements are sufficiently challenging that it is enjoyable to do them well and to develop
improved techniques.  However, for now, the overall results are sufficiently puzzling that there is plenty more to do.

\vspace*{-5mm}



%
%

\begin{theacknowledgments}
  I express my sincere gratitude to the organizers of both ISMD and WPCF.  Both actually succeeded
  in what is often just a stated but unattained goal of such workshops: to achieve meaningful intellectual
  cross-pollenation between nuclear and particle physicists.  I express
  special congratulations to the initiators of WPCF, which itself sprang out of the earlier Warsaw
  meetings, for establishing a timely and most promising workshop series.
\end{theacknowledgments}



\bibliographystyle{aipproc}   

\bibliography{Femtoscopy}

\begin{thebibliography}{56}
\expandafter\ifx\csname natexlab\endcsname\relax\def\natexlab#1{#1}\fi
\providecommand{\enquote}[1]{``#1''}
\expandafter\ifx\csname url\endcsname\relax
  \def\url#1{\texttt{#1}}\fi
\expandafter\ifx\csname urlprefix\endcsname\relax\def\urlprefix{URL }\fi
\providecommand{\eprint}[2][]{\url{#2}}

\bibitem[Lednicky(2005{\natexlab{a}})]{Lednicky:2005af}
R.~Lednicky  (2005{\natexlab{a}}), \eprint{nucl-th/0510020}.

\bibitem[Gyulassy and McLerran(2005)]{Gyulassy:2004zy}
M.~Gyulassy, and L.~McLerran, \emph{Nucl. Phys.} \textbf{A750}, 30--63 (2005),
  \eprint{nucl-th/0405013}.

\bibitem[Rischke and Gyulassy(1996)]{Rischke:1996em}
D.~H. Rischke, and M.~Gyulassy, \emph{Nucl. Phys.} \textbf{A608}, 479--512
  (1996), \eprint{nucl-th/9606039}.

\bibitem[Huovinen et~al.(2001)]{Huovinen:2001cy}
P.~Huovinen, P.~F. Kolb, U.~W. Heinz, P.~V. Ruuskanen, and S.~A. Voloshin,
  \emph{Phys. Lett.} \textbf{B503}, 58--64 (2001), \eprint{hep-ph/0101136}.

\bibitem[Ollitrault(1998)]{Ollitrault:1997vz}
J.-Y. Ollitrault, \emph{Nucl. Phys.} \textbf{A638}, 195--206 (1998),
  \eprint{nucl-ex/9802005}.

\bibitem[Adams et~al.(2004{\natexlab{a}})]{Adams:2004wz}
J.~Adams, et~al., \emph{Phys. Rev. Lett.} \textbf{93}, 252301
  (2004{\natexlab{a}}), \eprint{nucl-ex/0407007}.

\bibitem[Cole(2005)]{ColeQM05}
B.~Cole, \emph{plenary presentation at Quark Matter 2005, Budapest}  (2005).

\bibitem[Fries et~al.(2003)]{Fries:2003vb}
R.~J. Fries, B.~Muller, C.~Nonaka, and S.~A. Bass, \emph{Phys. Rev. Lett.}
  \textbf{90}, 202303 (2003), \eprint{nucl-th/0301087}.

\bibitem[Molnar(2005)]{Molnar:2004zj}
D.~Molnar, \emph{Acta Phys. Hung.} \textbf{A22}, 271--279 (2005),
  \eprint{nucl-th/0406066}.

\bibitem[Csorgo(2005)]{Csorgo:2005gd}
T.~Csorgo  (2005), \eprint{nucl-th/0505019}.

\bibitem[Padula(2005)]{Padula:2004ba}
S.~S. Padula, \emph{Braz. J. Phys.} \textbf{35}, 70--99 (2005),
  \eprint{nucl-th/0412103}.

\bibitem[Lisa et~al.(2005)]{Lisa:2005dd}
M.~A. Lisa, S.~Pratt, R.~Soltz, and U.~Wiedemann, \emph{Ann. Rev. Nucl. Part.
  Sci.} \textbf{55}, 311 (2005), \eprint{nucl-ex/0505014}.

\bibitem[Heinz et~al.(2002)]{Heinz:2002au}
U.~W. Heinz, A.~Hummel, M.~A. Lisa, and U.~A. Wiedemann, \emph{Phys. Rev.}
  \textbf{C66}, 044903 (2002), \eprint{nucl-th/0207003}.

\bibitem[Bartke(1986)]{Bartke:1986mj}
J.~Bartke, \emph{Phys. Lett.} \textbf{B174}, 32--35 (1986).

\bibitem[Alexander(2003)]{Alexander:2003ug}
G.~Alexander, \emph{Rept. Prog. Phys.} \textbf{66}, 481--522 (2003),
  \eprint{hep-ph/0302130}.

\bibitem[Humanic et~al.(1988)]{Humanic:1988ny}
T.~J. Humanic, et~al., \emph{Z. Phys.} \textbf{C38}, 79--84 (1988).

\bibitem[Harris and Muller(1996)]{Harris:1996zx}
J.~W. Harris, and B.~Muller, \emph{Ann. Rev. Nucl. Part. Sci.} \textbf{46},
  71--107 (1996), \eprint{hep-ph/9602235}.

\bibitem[Bass et~al.(1999)]{Bass:1999zq}
S.~A. Bass, et~al., \emph{Nucl. Phys.} \textbf{A661}, 205--260 (1999),
  \eprint{nucl-th/9907090}.

\bibitem[Gyulassy(2004)]{Gyulassy:2004jj}
M.~Gyulassy, \emph{J. Phys.} \textbf{G30}, S911--S918 (2004).

\bibitem[Bialas et~al.(1976)]{Bialas:1976ed}
A.~Bialas, M.~Bleszynski, and W.~Czyz, \emph{Nucl. Phys.} \textbf{B111}, 461
  (1976).

\bibitem[Adamova et~al.(2003)]{Adamova:2002ff}
D.~Adamova, et~al., \emph{Phys. Rev. Lett.} \textbf{90}, 022301 (2003),
  \eprint{nucl-ex/0207008}.

\bibitem[Back et~al.(2005)]{Back:2004je}
B.~B. Back, et~al., \emph{Nucl. Phys.} \textbf{A757}, 28--101 (2005),
  \eprint{nucl-ex/0410022}.

\bibitem[Adams et~al.(2005)]{Adams:2004yc}
J.~Adams, et~al., \emph{Phys. Rev.} \textbf{C71}, 044906 (2005),
  \eprint{nucl-ex/0411036}.

\bibitem[Chajecki(2005{\natexlab{a}})]{Zibi:WPCF}
Z.~Chajecki, \emph{contribution to these proceedings}  (2005{\natexlab{a}}).

\bibitem[Chajecki(2005{\natexlab{b}})]{Chajecki:2005zw}
Z.~Chajecki  (2005{\natexlab{b}}), \eprint{nucl-ex/0510014}.

\bibitem[Retiere and Lisa(2004)]{Retiere:2003kf}
F.~Retiere, and M.~A. Lisa, \emph{Phys. Rev.} \textbf{C70}, 044907 (2004),
  \eprint{nucl-th/0312024}.

\bibitem[Adams et~al.(2004{\natexlab{b}})]{Adams:2003ra}
J.~Adams, et~al., \emph{Phys. Rev. Lett.} \textbf{93}, 012301
  (2004{\natexlab{b}}), \eprint{nucl-ex/0312009}.

\bibitem[Adams et~al.(2004{\natexlab{c}})]{Adams:2003zg}
J.~Adams, et~al., \emph{Phys. Rev. Lett.} \textbf{92}, 062301
  (2004{\natexlab{c}}), \eprint{nucl-ex/0310029}.

\bibitem[Lisa(2004)]{Lisa:2003ze}
M.~A. Lisa, \emph{Acta Phys. Polon.} \textbf{B35}, 37--46 (2004),
  \eprint{nucl-ex/0312012}.

\bibitem[Teaney et~al.(2001)]{Teaney:2001av}
D.~Teaney, J.~Lauret, and E.~V. Shuryak  (2001), \eprint{nucl-th/0110037}.

\bibitem[Heinz and Kolb(2002{\natexlab{a}})]{Heinz:2002sq}
U.~W. Heinz, and P.~F. Kolb, \emph{Phys. Lett.} \textbf{B542}, 216--222
  (2002{\natexlab{a}}), \eprint{hep-ph/0206278}.

\bibitem[Yano and Koonin(1978)]{Yano:1978gk}
F.~B. Yano, and S.~E. Koonin, \emph{Phys. Lett.} \textbf{B78}, 556--559 (1978).

\bibitem[Wu et~al.(1998)]{Wu:1996wk}
Y.~F. Wu, U.~W. Heinz, B.~Tomasik, and U.~A. Wiedemann, \emph{Eur. Phys. J.}
  \textbf{C1}, 599--617 (1998), \eprint{nucl-th/9607044}.

\bibitem[Retiere(2004)]{Retiere:2004wa}
F.~Retiere, \emph{J. Phys.} \textbf{G30}, S827--S834 (2004),
  \eprint{nucl-ex/0405024}.

\bibitem[Akkelin and Sinyukov(1995)]{Akkelin:1995gh}
S.~V. Akkelin, and Y.~M. Sinyukov, \emph{Phys. Lett.} \textbf{B356}, 525--530
  (1995).

\bibitem[Tomasik(2005)]{Tomasik:2005ny}
B.~Tomasik  (2005), \eprint{nucl-th/0509100}.

\bibitem[Chajecki et~al.(2005)]{Chajecki:2005qm}
Z.~Chajecki, T.~D. Gutierrez, M.~A. Lisa, and M.~Lopez-Noriega  (2005),
  \eprint{nucl-ex/0505009}.

\bibitem[Chaloupka(2005)]{Chaloupka:WPCF}
P.~Chaloupka, \emph{contribution to these proceedings}  (2005).

\bibitem[Adams(2005)]{Adams:2005ws}
J.~Adams  (2005), \eprint{nucl-ex/0511003}.

\bibitem[Kisiel(2005)]{Adam:WPCF}
A.~Kisiel, \emph{contribution to these proceedings}  (2005).

\bibitem[Gos(2005)]{Hanna:WPCF}
H.~Gos, \emph{contribution to these proceedings}  (2005).

\bibitem[Heffner(2004)]{Heffner:2004js}
M.~Heffner, \emph{J. Phys.} \textbf{G30}, S1043--S1047 (2004).

\bibitem[Bekele(2004)]{Bekele:2004ci}
S.~Bekele, \emph{J. Phys.} \textbf{G30}, S229--S234 (2004).

\bibitem[Adams et~al.(2003)]{Adams:2003qa}
J.~Adams, et~al., \emph{Phys. Rev. Lett.} \textbf{91}, 262302 (2003),
  \eprint{nucl-ex/0307025}.

\bibitem[Adler et~al.(2004)]{Adler:2004rq}
S.~S. Adler, et~al., \emph{Phys. Rev. Lett.} \textbf{93}, 152302 (2004),
  \eprint{nucl-ex/0401003}.

\bibitem[Back et~al.(2004)]{Back:2004ug}
B.~B. Back, et~al.  (2004), \eprint{nucl-ex/0409001}.

\bibitem[Lednicky(2001)]{Lednicky:2001qv}
R.~Lednicky  (2001), \eprint{nucl-th/0112011}.

\bibitem[Lednicky(2005{\natexlab{b}})]{Lednicky:WPCF}
R.~Lednicky, \emph{contribution to these proceedings}  (2005{\natexlab{b}}).

\bibitem[Heinz and Kolb(2002{\natexlab{b}})]{Heinz:2002un}
U.~W. Heinz, and P.~F. Kolb  (2002{\natexlab{b}}), \eprint{hep-ph/0204061}.

\bibitem[Roland et~al.(2005)]{Roland:2005ei}
G.~Roland, et~al.  (2005), \eprint{nucl-ex/0510042}.

\bibitem[Caines(2005)]{caines:QM05}
H.~Caines, \emph{poster presentation at Quark Matter 2005, Budapest}  (2005).

\bibitem[Bystersky and Retiere(2005)]{Retiere:WPCF}
M.~Bystersky, and F.~Retiere, \emph{contribution to these proceedings}  (2005).

\bibitem[Danielewicz and Pratt(2005)]{Danielewicz:2005qh}
P.~Danielewicz, and S.~Pratt  (2005), \eprint{nucl-th/0501003}.

\bibitem[Danielewicz(2005)]{Danielewicz:WPCF}
P.~Danielewicz, \emph{contribution to these proceedings}  (2005).

\bibitem[Brown(2005)]{Brown:WPCF}
D.~Brown, \emph{contribution to these proceedings}  (2005).

\bibitem[Chung(2005)]{Chung:WPCF}
P.~Chung, \emph{contribution to these proceedings}  (2005).

\end{thebibliography}

\end{document}